\documentstyle[12pt,amsmath,amssymb,exscale,relsize,graphicx]{article}

\date{\today}

\newcommand{\Ccal}{{\mathcal{C}}}

\newcommand{\Scal}{{\mathcal{S}}}

\newenvironment{pf}{\medskip \par \noindent {\it Proof.}\ }
{\hfill \P \bigskip \par}

\newcommand{\beq}{\begin{equation}}
\newcommand{\eeq}{\end{equation}}

\newtheorem{thm}{Theorem}

\newtheorem{alg}{Algorithm}
\newtheorem{exmp}{Example}

\title{FastLSU - A more practical approach for the Benjamini-Hochberg FDR controlling procedure for huge-scale testing problems.}
\author{Vered Madar$^*$
Department of Biostatistics,\\ University of North Carolina, Chapel Hill,
	   NC 27599, USA,\\ and\\
Sandra  Batista$^{**}$
Department of Computer Science,\\ Princeton University, Princeton,
	   NJ 08540, USA\\ }

\begin{document} 
\maketitle
\begin{abstract}
We address a common problem in large-scale data analysis, and especially the field of genetics, the huge-scale testing problem, where millions to billions of hypotheses are tested together creating a computational challenge to perform multiple hypotheses testing procedures. 
As a solution we propose an alternative algorithm to the well used Linear Step Up procedure of Benjamini and Hochberg (1995). 
Our algorithm requires linear time and does not require any p-value ordering. It permits separating huge-scale testing problems arbitrarily into computationally feasible sets or \textit{chunks}. Results from the \textit{chunks} are combined by our algorithm to produce the same results as the controlling procedure on the entire set of tests, thus controlling the global false discovery rate even when p-values are arbitrarily divided. The practical memory usage may also be determined arbitrarily by the size of available memory. 
\end{abstract}

\textbf{keywords:} Adjusted p-values, chunks of p-values, expression quantitative trait loci, false discovery rate, false discovery proportion, family-wise error rate, huge-scale testing problems, large-scale testing problems, linear step-up;

\section{The problem of huge-scale data and separating hypotheses tests}
In many fields the substantially increased scale of data available has resulted  in a significant increase in size of multiple hypotheses testing problems. In genetics in particular, new advances have led to typical GWAS studies consisting of $10^6$ SNPs~\cite{Hindroff} 
and eQTL studies usually consisting of $10^9$ tests~\cite{XiaEtAl2012,WrightEtAl2014}. These testing problems are of \textit{huge-scale} as opposed to \textit{large-scale} used by Efron~\cite{Efron2004} to describe studies consisting hundreds to thousands hypotheses. 
These huge-scale multiple hypotheses testing problems create numerous computational challenges when many tests, say of the order $10^6$, are performed with all of the p-values of more or less equal importance.
As a result some simpler testing procedures such as rigid p-value thresholds may be used that sacrifice power and correctness.

Alternatively tests may be separated or chunked into smaller sets or chunks that are more computationally feasible.  
Efron~\cite{Efron2008} notes that the problem of separating hypotheses tests has not received great attention 
and warns of some pitfalls in chunking p-values, but focuses on grouping tests that share a  biological property rather than arbitrary, computationally feasible chunks.
Cai and Sun~\cite{CaiSun2009} and later Benjamini and Bogomolov~\cite{BenjaminiBogomolov2014} propose alternative solutions to Efron's grouping problem  
but do not address the problem of arbitrary, computationally feasible chunking.      
We confront the computationally feasible chunking problem for the Benjamini-Hochberg false discovery rate~\cite{BenjaminiHochberg1995}.
We show on data from Stranger's HapMap study~\cite{Stranger2007,XiaEtAl2012} that if results from separate tests are not combined correctly,
there is considerable inflation of type I error, offer an explication for this occurrence, and propose our algorithm as a solution.

Consider a huge-scale testing problem of size $m$ where our goal is to select exactly $R \ge 0$ significant tests. 
Of the $R$ significant discoveries, exactly $V \ge 0$ tests will be false discoveries (i.e., truly non-significant tests that are declared significant). 
A common approach in multiple comparison testing is to control the family-wise error rate, $FWER = \Pr(V\ge 1)$, the probability to make at least one false discovery. Alternatives are to control the false discovery rate ($FDR$), the averaged value of the false discovery proportion, $FDP = V/min(R,1)$, the proportion of truly false tests among the significant $R$, or the positive FDR, $pFDR = E(V/R|R>0)$, the average of the $FDP$ when significant tests are selected. For a further discussion about $FWER$, $FDR$,  $pFDR$, and their variations refer to Farcomeni~\cite{Farcomeni2009}.

In a huge-scale testing, when the $m$ p-values are partitioned into chunks,
it is challenging to control any of the above error rates ($FWER$, $pFDR$ or $FDR$) over the entire collection of $m$ p-values. 
Controlling these error rates on a per chunk basis, if not done correctly, may interfere with the overall results by introducing more false discoveries. 
Consider for instance an example of $FWER$ control, the Bonferroni approach that for a chunk of size $m$ collects all p-values less than $\alpha/m$ 
Applying Bonferroni in chunks of size $m_i$ will tend to select more significant results than applying it over the entire set of $m = \sum m_i$ p-values since $\alpha/m$ is less than $\alpha/m_i$. 
In the case of $FWER$ control using a fixed bound of $\alpha/m$ for all the chunks is preferred. 
A stricter constant cut-off on all sets of tests is suggested by Dudbridge and Gusnanto for GWAS~\cite{DudbridgeGusnanto2008}. 

It is preferable to control the false discovery proportion rather than the number of false positives for a huge-scale testing problem. 
Therefore the $FDR$ or the $pFDR$ approaches are favored and both tend to offer larger, more powerful sets of results than those that might be offered by the conservative $FWER$ control. 

\subsection{The Benjamini-Hochberg linear step up procedure for controlling the false discovery rate}\label{sec:explainLSU}

The Benjamini-Hochberg Linear Step Up(LSU) procedure~\cite{BenjaminiHochberg1995} is designed to control the False Discovery Rate, the expected value of the false discovery proportion, i.e.,
 $FDR = E(FDP)$. The $FDR$ is at most equal to $FWER$ and $FDR = FWER$ when all tests are non-significant in truth. 
As a result for huge scale multiple hypotheses tests of equal importance, controlling a proportion of false discoveries, especially on the average, has  increased power over procedures that control the FWER such as Bonferroni or a rigid 
cut-off bound such as $5\times 10^{-8}$ suggested for GWAS~\cite{DudbridgeGusnanto2008}. The larger the multiple hypotheses testing problem is the more powerful the LSU is over procedures that control the $FWER$.

While the LSU procedure~\cite{BenjaminiHochberg1995} is still one of the most cited procedures, its application had required sorting all p-values in decreasing order to look for the largest p-value that satisfies a simple condition. 
In face of a huge scale testing problem rather than apply LSU, some researchers had preferred to use harsh p-value cut-offs as mentioned above or to divide their huge-scale set of tests into computationally feasible smaller chunks 
and apply the multiple hypotheses testing procedure on each chunk selecting as the final significant results the union of the results in each of the chunks. 
Efron~\cite{Efron2008} warns of the danger in such aggregation from the perspective of $pFDR$, pointing out that some chunks might have a larger proportion of significant results than others, and aggregating the significant results can yield misleading estimates. 
Moreover different chunking of tests might yield different sets of significant results. Analysis done by different groups of populations or chromosomes may not give the same number of significant tests as analysis that is applied on equal sized subsets; for example, it is well known that chromosome 6 has a higher proportion of significant HLA SNPs than other chromosomes.  We shall show that sorting the p-values, arbitrary thresholds, and arbitrary aggregation of results are not necessary and do not
improve computational space or time efficiency compared to our algorithm.

\subsection{A Faster Algorithm for LSU}
Our alternative algorithm to the LSU, \textbf{FastLSU}, performs linear scans instead of sorting p-values, but takes into account the overall size of the testing problem. FastLSU tiles the LSU procedure to give one global set of results
 that does not differ from applying LSU to the entire set of tests. Our algorithm addresses the same objective function as the original. Our approach is provably faster than the conventional approach that relies on sorting p-values.
It may also be used on arbitrary chunks of arbitrary size with an arbitrary space constraint in order to return the same set of significant results as those from applying LSU to the entire set of tests.

In the following section we address the difference between grouping and chunking tests and the difficulty in arbitrarily chunking tests by giving
examples of inflation of type I error.  In section~\ref{sec:FastLSU_algs} we present FastLSU on a single set of tests and prove its equivalence to LSU, time efficiency, and space efficiency. We present FastLSU on arbitrary chunks and also show its correctness and
 efficiency  in Section~\ref{sec:FastLSU_A2}. We offer suggestions for finalizing the report of significant tests in Section~\ref{sec:app_tips} and  conclude with discussion in Section~\ref{sec:discussion}. 
Code for an implementation of the algorithm are given in R and SAS in the Appendix~\ref{app:proofs}. 

\section{An exercise of three HapMap groups and their chunking for Stranger's cis-eQTL study}\label{sec:badchunks}
FastLSU controls the global $FDR$, not a family-based bound or a group-based bound.
Our final results are not affected by the procedure of separating tests.
This is an important distinction because chunking is arbitrary and based primarily on computational efficiency without any consideration for relationships between the hypotheses being tested. 
This is in contrast to group-testing procedures where for example, hypotheses may be grouped based on experimental knowledge such as all the
tests from the same chromosome, the back and front parts of the brain~\cite{Efron2008}, or different population groups in HapMap~\cite{Stranger2007}. 
In Section~\ref{sec:groupie} we will show how the FastLSU algorithm can even improve the efficiency of a group-based controlling procedure. 
FastLSU is particularly suited to the current applications in genetics because we typically seek the significant set of SNPs or genes, and the family structure is usually less importance than managing the computational burden. 
Genetic family structures typically do not require any correction for family selection because each genetic family tends to contribute significant results of its own. (This is the case for the following example of 3 families of HapMap.) 
For this reason in the remainder of this section, we will give motivating examples of applying FastLSU on the three groups of approximately $14\times10^6$ cis-eQTLs of Stranger's HapMap study~\cite{Stranger2007,XiaEtAl2012} in order
to demonstrate the problems with arbitrary chunking without combining the results as FastLSU does.

\begin{figure}[H]
\centering
\includegraphics[width=1\columnwidth]{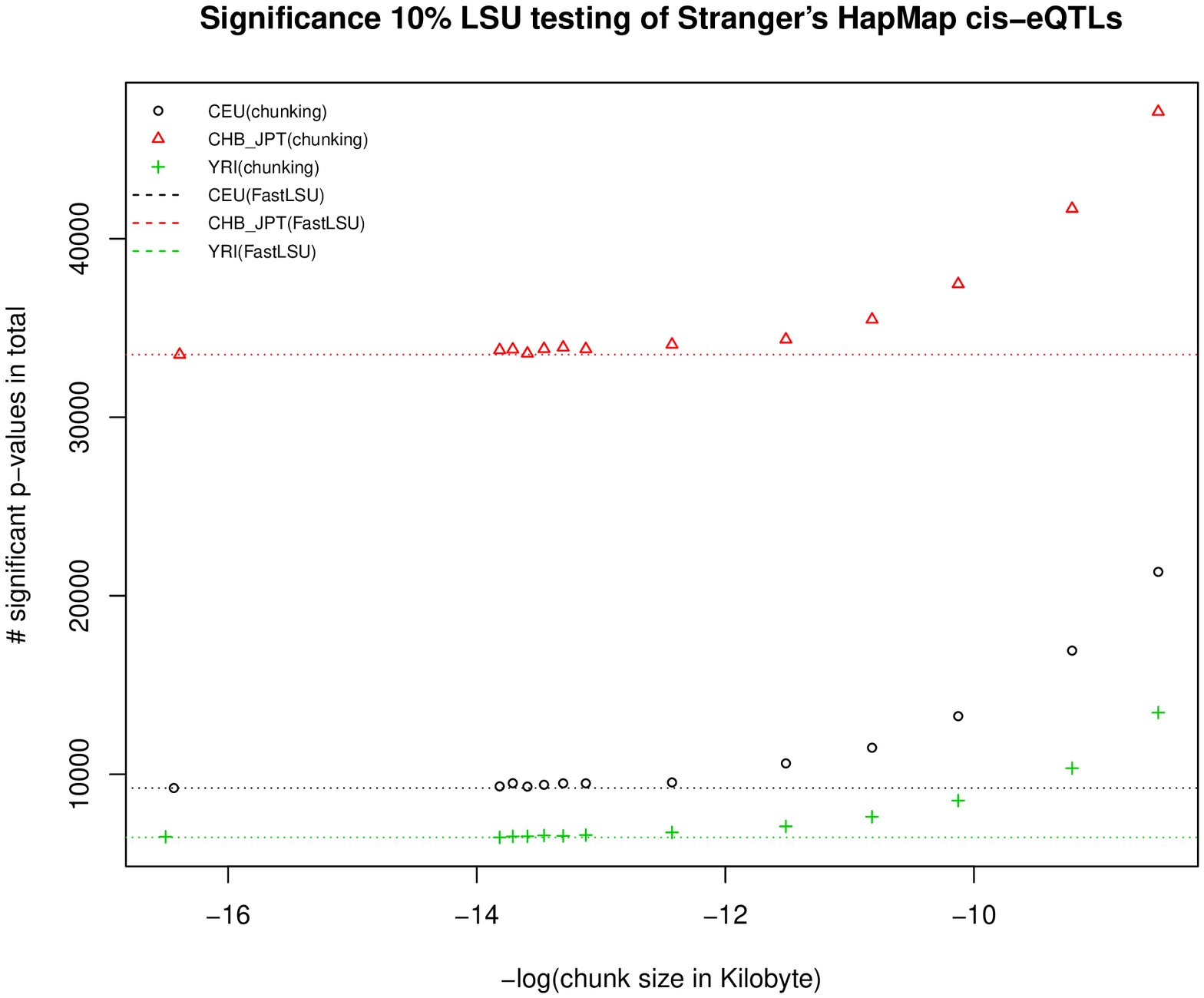}
\caption{Varying chunk size over Stranger's HapMap populations. Solid lines represent results of applying the LSU under $10\%$ 
Dashed lines show the consistent result of applying the FastLSU as in Algorithm~\ref{alg:2chunks}. }
\label{fig:chunking_exe}
\end{figure}

Stranger~\cite{Stranger2007} presents an eQTL study over 4 HapMap population samples: $30$ Central Europeans(CEU), $45$ Chinese (CHB), $45$ Japanese(JPT) and $30$ trios from Nigerians(YRI). To increase power each group is analyzed separately. 
We follow the recommendations of the SeeQTL website~\cite{XiaEtAl2012} and consider the CHB and JPT together. We define cis-eQTLs as within $1$ Mb upstream or downstream of a gene.
 Each populations has a set of approximately $14$ million
p-values that were split into chunks in sizes: 
$1M$, $900K$, $800K$, $700K$, $600K$, $500K$, $250K$, $100K$, $50K$, $25K$ to $10K$ and $5K$.
We contrast the differences in the number of significant results when the results are combined by taking the union of the results of LSU on each
chunk versus those selected by FastLSU Algorithm~\ref{alg:2chunks} in Figure~\ref{fig:chunking_exe}.

Applying the FastLSU yields $9,228$ significant results for CEU, $6,497$ for YRI and $33,507$ for the CHB \& JPT group. Most notable is that the results for 
FastLSU Algorithm~\ref{alg:2chunks} do not change across the chunk sizes whereas the alternative of taking the union of the
results on each chunk increases not only the number of significant tests selected, but also the maximal p-value reported as the chunk size decreases.  
For example, applying FastLSU at level $10\%$ for the 
complete chunk of approximately $14M$ p-values yields $9,228$ discoveries for the CEU; however, applying LSU on $1,370$ chunks of $10K$ yields $16,925$ significances which is equivalent to an overall FDR control of $23.75\%$. 
If we assess the magnitude of the significance level required to get the same number of significances for $10K$ sized chunks in the above exercise, we see that the CHB-JPT group $41,587$ requires $\alpha = 15.58\%$ 
and for the YRI group $\alpha = 21.1\%$ will give $10,330$ significant p-values. This implies that performing the analysis using  $1,300$ to $1,400$ chunks of size 10K rather than a single chunk inflates the type I error by $50\%$.  
While this is compelling
experimental evidence, a more compelling explanation is that the union of partial orderings on subsets of a set is in general not equal to the partial ordering on the entire set. Moreover when the size of the chunks decreases,
the significance interval is being divided into larger sub-intervals more likely each containing more p-values spanning a greater range of values, so that the set of selected significant p-values  will
more likely contain a greater maximal p-value. The max p-value we obtain for the case of single chunk is $6.7\times 10^{-5}$ for CEU, $4.4\times 10^{-5}$ for YRI, and $2.5\times 10^{-4}$ for CHB-JPT. For chunks of the size $100K$ (about $13$ to $14$ chunks), the max p-values are slightly higher especially for the CHB-JPT group: $1.7\times 10^{-4}$ for CEU, $3.5\times 10^{-4}$ for YRI and $1.2\times 10^{-3}$ for CHB-JPT. 
For chunks of $10K$ size (about $130$ to $140$ chunks) the maximal p-values are considerably higher still: $0.0063$ for CEU, $0.0034$ for YRI and $0.011$ for CHB-JPT.  As we mentioned, there is no need to apply any correction for group
of family selection as required by Benjamini and Bogomolov~\cite{BenjaminiBogomolov2014} since all three groups in the HapMap example give significant results.

\section{The FastLSU algorithm}\label{sec:FastLSU_algs}
The usual way to apply the LSU~\cite{BenjaminiHochberg1995} at level $0 \le \alpha \le 1$ is to sort the p-values in ascending order, 
$p_{(m)} \ge p_{(m-1)} \ge \cdots \ge p_{(1)} $. 
Starting from the largest p-value to the smaller we need to look for the largest p-value that satisfy $p_{(k)} < k\alpha/m$.
One may view the LSU algorithm~\cite{BenjaminiHochberg1995} as a search algorithm for 
\begin{equation}\label{eq:objLSU}
   r =  argmax\{i : \: p_{(i)} < i \alpha / m\}.
\end{equation}
This observation motivates the following Fast Linear Step UP (FastLSU) algorithm that controls the FDR at level $\alpha$, for $0<\alpha<1$:
\begin{alg}\label{alg:1chunk}
\textbf{FastLSU for a single batch - without sorting the p-values}
\begin{itemize}
\item[\textbf{Step 1.}] Start with $r_0=m$ and count all p-values $< r_0 \alpha / m$; let $r_1$ be their count.
\item[\textbf{Step k.}] Do for $k = 2, \ldots, m$  Count all p-values $< r_{k-1} \alpha / m$; let $r_k$ be their count.
\item[\textbf{End.}] Repeat step $k$ for $k = k+ 1$ until $r_k = r_{k+1}$ or $k = m$.
\end{itemize}
\end{alg}

\subsection{The equivalence to LSU and computational efficiency}
\begin{thm}\label{thm:Algorithm1}
For a significance level $\alpha$, the Algorithm~\ref{alg:1chunk} maximizes the same objective function~\ref{eq:objLSU} that is used by the LSU.
Therefore, the FastLSU controls the $FDR$ at level $\alpha$ and gives the same selected set of significant results as would be obtained by applying the Benjamini-Hochberg~\cite{BenjaminiHochberg1995} LSU $FDR$ controlling procedure. 
\end{thm}  

\begin{pf}
Consider a batch of p-values which we will denote as $\Ccal$. Let $0 \le t \le 1$ and define $\Scal(t:\Ccal) = \{p<t \,:\: p \mbox{ is p-value} \in \Ccal\}$ to be the number of p-values from $\Ccal$ smaller than $t$. 
The set $\Scal\left(\frac{i\alpha}{m}\right)$ consists of all p-values from $\Ccal$ that are smaller than $\frac{i\alpha}{m}$. Hence, for the case of a single batch, it is possible to verify that the search for 
\begin{equation}\label{argmax:alg1}
    r = argmax\{i : \: i=\#\Scal\left( i\alpha/m\right) \} 
\end{equation}
 is equivalent to looking for the largest $r$th p-value satisfying $p_{(r)} \le \frac{r\alpha}{m}$, as requested by (\ref{eq:objLSU}).
\end{pf}

\begin{thm}\label{thm:cmp_Om}
 The FastLSU Algorithm~\ref{alg:1chunk} requires $O(m)$ time and $O(m)$ space where $m$ is the number of p-values to be considered.
\end{thm}
The proof to Theorem~\ref{thm:cmp_Om} is given in the Appendix~\ref{app:proofs} with accompanying pseudocode for procedural language implementations. 

\begin{exmp}\label{ex:exmp1}
To demonstrate this equivalence we apply FastLSU on the example appearing in the original 1995 FDR paper~\cite{BenjaminiHochberg1995} with $\alpha = 0.05$. Instead of sorting all $15$ p-values we apply Algorithm~\ref{alg:1chunk}
using $5$ linear scans and find $4$ significant p-values. Consider the $15$ p-values from the example given at LSU paper~\cite{BenjaminiHochberg1995}:
\[
0.6528,0.7590,0.0298,0.4262,0.0459,0.0278,0.0001,0.0019,
\]
\[
0.0004,0.0201,1.0000,0.5719,0.3240,0.0095,0.0344
\]
Scanning the p-values, for the first time, we find $9$ p-values $< 0.05$,
the  second $r_2=7$ p-values $< 9 \cdot 0.05/15$, and the third $r_3 = 5$ p-values $< 7\cdot 0.05/15$. 
The fifth scan finds exactly $r_4 = 4$ p-values that satisfy $< 4\cdot 0.05/15$.
The selected p-values are $0.0001,0.0019,0.0004,0.0095$, and the reader can check that these $4$ p-values are exactly the ones selected using the original LSU p-value sorting algorithm.
\end{exmp}

\begin{exmp}[Example of huge-scale run]\label{ex:exmp2}
We apply the Algorithm~\ref{alg:1chunk} over the group of $13,755,172$ Central Europeans(CEU) HapMap p-values of Stranger~\cite{Stranger2007} that was described in Section~\label{sec:badchunks}. For this example, the FastLSU makes $11$ linear scans to achieve the $9,228$ significant p-values, and the running time is of several seconds.
\end{exmp}
\subsection{Applying FastLSU over families or groups of p-values of equal relevance}\label{sec:groupie}
Efron~\cite{Efron2008} distinguishes between the groups of imaging p-values that  arise from the front brain and the back brain, and develops an empirical Bayes set-up to combine the significant p-values from the two families.
Benjamini and Bogomolov~\cite{BenjaminiBogomolov2014} extended Efron's idea of groups into the voxel families of MRI where each set of p-values
from a specific voxel of a certain location are treated as a separate, homogeneous group of relevance.
For the first step each voxel group is analyzed by applying the LSU at level $\alpha$. For the second step the number of significant groups is collected and LSU is performed again with another significance level $\alpha^*$ 
to correct for the selection of groups. If, for example, $S$ groups out of $G$ are shown to have at least one significant p-value for LSU of level $\alpha$, $\alpha^*$ is set to $\frac{S}{G}\alpha$ for each of the groups~\cite{BenjaminiBogomolov2014}. 

Given the equivalence between LSU and FastLSU, FastLSU may be used for each step in the approach of Benjamini and Bogomolov. The second step need only be done for groups that have at least one significant result. 
 Since only significant p-values for each group need to be considered,  FastLSU may be applied to each group by using $\alpha^{**} = \frac{S r_g(\alpha)}{G m_g}\alpha$  where each group 
$g$ ($g =1, 2,\ldots,G$) contains $m_g$ p-values of which $r_g(\alpha)$ were declared significant of level $\alpha$. Since FastLSU achieves efficiency  by the simple recalling and correction of the constants
$m$ (or $m_g$) for the number of tests, a similar approach also works for Storey-Efron's positive FDR approach~\cite{Efron2004, Storey2002}. 
To be more precise, the complete grid of $\lambda$ values should be collected in each step and then the q-values should be corrected accordingly. 

\section{Applying LSU on arbitrary chunks of p-values}\label{sec:FastLSU_A2}
Algorithm~\ref{alg:1chunk} may be extended for when p-values are divided into an arbitrary number of chunks of arbitrary size. The size of the chunks may be determined by practical memory constraints
or any other criteria. The iterative steps are applied on each chunk. The significant results from each chunk are combined using another iterative step to form the final set of significant results.

\begin{alg}\label{alg:2chunks}\textbf{FastLSU for the case of two or more chunks of p-values}
Given a set of $m$ tests and p-values divided into $n$ separate chunks of sizes $m_{c_i}$ p-values for $i=1,...,n$ such that $m = \sum_{i=1}^{n} m_{c_i}$:
\begin{itemize}
\item{\textbf{Step $1$.}} On each of the chunks, count the number of p-values less than $\alpha$. Denote this count as $r_{c_i}^{(1)}$.
\item{\textbf{Step $k+1$.}} On each of the chunks (starting from $k=1$) count the number 
of p-values less than $( \sum_{i=1}^{n} r_{c_i}^{(k)} ) \alpha / m$, and let $r_{c_i}^{(k+1)}$ be the count for the chunk at this step. 
\item{\textbf{Repeat Step $k+1$}} until $k+1 = m$ or $\sum_{i=1}^{n} r_{c_i}^{(k+1)} = \sum_{i=1}^{n} r_{c_i}^{(k)}$. Mark these p-values as significant under LSU. 
\end{itemize}
\end{alg}
An example of Algorithm~\ref{alg:2chunks} applied to the example appearing in the original 1995 FDR paper~\cite{BenjaminiHochberg1995} is given in the Appendix.

There are many ways one can alter Algorithm~\ref{alg:2chunks}, either by flagging or dropping p-values that do not satisfy the condition in the iterative step that preserves the LSU. 
Alternatively Algorithm~\ref{alg:1chunk} may be applied to each chunk independently starting with the initial value for $r_0 = m - m_{c_i} + r_{c_1}^{(1)}$ and still tiling by the total number 
of tests, $m$. (The advantage of this is that this may be done in parallel if desired.) The results of the chunks may then be combined into a single set, if space permits, and Algorithm 1 applied to this set to return the final set of significant results. If space does not permit the results of the
chunks to be combined, they may be combined up to the space limit. Then the iterative step of Algorithm~\ref{alg:2chunks} above may be applied to the resultant chunks starting by checking for  p-values less than  $m' \alpha /m$ where $m'$
is the size of the union of the results.

\begin{thm}\label{thm:Algorithm2}
For a significance level $\alpha$, and a collection of $c$ chunks the Algorithm~\ref{alg:2chunks} gives the same selected set of significant results as would be obtained by applying the Benjamini-Hochberg~\cite{BenjaminiHochberg1995} LSU $FDR$ controlling procedure over the single chunk consisting the union of all p-values  from the $c$ chunks. 
\end{thm}  

\begin{pf}[Proof for Algorithm~\ref{alg:2chunks}]
Suppose we have exactly $c$ disjoint chunks of p-values $\Ccal_1,\Ccal_2, \ldots, \Ccal_c$, with  sizes  $m_{\Ccal_i}$ ($i=1,2, \ldots,c$) such that $\Ccal = \cup_{i=1}^c \Ccal_i$ and  $|\Ccal| = m=\sum_{i=1}^c m_{\Ccal_i}$ p-values. Let $r_\Ccal$ be the number of selected significant p-values that satisfying the LSU objective (\ref{eq:objLSU}):
\begin{equation}\label{eq:all}
    r_\Ccal  
         = argmax\left\{ r :\  r = \#\Scal\left(\frac{r \alpha}{m}\, :\Ccal\right)\right\}.
\end{equation}

Algorithm~\ref{alg:2chunks} can be applied to each $c$ chunks and the last step of Algorithm~\ref{alg:2chunks} is to finalize the selection.
Accordingly for chunk $\Ccal_i$ we search for the largest $r_{\Ccal_i}^*$ that satisfy the objective:
 \begin{equation}\label{eq:alg2}
     r_{\Ccal_i}^*  
         = argmax\left\{ r :\  r = \#\Scal\left(\frac{(r+ m - m_{\Ccal_i})\alpha}{m}\, :\Ccal_i\right)\right\}.
\end{equation}
We claim that
\begin{equation}\label{ineq:rGcal}
      r_\Ccal \le  r_{\Ccal_i}^* + m - m_{\Ccal_i},\quad \mbox{for any } i=1, \ldots,c
\end{equation}
This implies that each of the p-values selected significant from the search for (\ref{eq:all}) within chunk $\Ccal_i$ must be selected while searching for argmax $r_{\Ccal_i}^*$ as in (\ref{eq:alg2}) since $p \le r_\Ccal \cdot \alpha/m$ implies $p \le (r_{\Ccal_i}^* + m - m_{\Ccal_i})\alpha/m$.

It is possible to verify that $p_{(k:\Ccal_i)}$ the $k$th p-value in chunk $\Ccal_i$ cannot be larger than, $p_{(k+ m - m_{\Ccal_i})}$, the $(k+ m - m_{\Ccal_i})$th p-value in the union $\Ccal$:
\begin{equation}\label{ineq:pv_orgenize}
p_{(k:\Ccal_i)} \le p_{(k+ m - m_{\Ccal_i})}
\qquad \mbox{ for all }k = 1, \ldots , m_{\Ccal_i}.
\end{equation}
In particular, by the condition (\ref{eq:alg2}),  let $p_{(r_{\Ccal_i}^*)}$ be the largest p-value in $\Ccal_i$ that satisfies 
$p_{(r_{\Ccal_i}^*:\Ccal_i)} <( r_{\Ccal_i}^*+m -m_{\Ccal_i})\alpha/m$. When $m_{\Ccal_i}> r_{\Ccal_i}^*$, it follows that
\begin{equation}\label{ineq:pvls_nonsig}
( r_{\Ccal_i}^*+ k + m -m_{\Ccal_i})\alpha/m < p_{(r_{\Ccal_i}^* + k:\Ccal_i)} \le p_{(r_{\Ccal_i}^* + k+ m - m_{\Ccal_i})}
\qquad \mbox{ for all } k = 1, \ldots , m_{\Ccal_i} - r_{\Ccal_i}^* .
\end{equation}
(When $m_{\Ccal_i}= r_{\Ccal_i}^*$ (\ref{ineq:rGcal}) holds trivially since at most the entire set of tests may be selected as significant.) To prove (\ref{ineq:rGcal}) by contradiction,
 let us compare between $p_{(r_\Ccal)}$ to $p_{(r_{\Ccal_i}^*+ k + m - m_{\Ccal_i})}$.
If we assume that $r_\Ccal > r_{\Ccal_i}^* + m - m_{\Ccal_i}$ , then we can define a positive integer $k =r_\Ccal - (r_{\Ccal_i}^* + m - m_{\Ccal_i})$ for which following the inequality in (\ref{ineq:pvls_nonsig}) holds, 
\[
 r_\Ccal\cdot \alpha/m = ( r_{\Ccal_i}^*+ k + m -m_{\Ccal_i})\alpha/m
< p_{(r_{\Ccal_i}^*+ k:\: \Ccal_i)} 
\le p_{(r_{\Ccal_i}^*+k+ m - m_{\Ccal_i})} = p_{(r_\Ccal)},
\]
However, this is in contradiction to the condition (\ref{eq:all}) that provides $p_{(r_\Ccal)} < r_\Ccal\cdot \alpha/m$.

In conclusion we show that applying FastLSU over the global union of p-values give the exact same selection of significant p-values. 
We search for $r_\Ccal^*$ over the union of the results on each chunk of size $\sum_{i=1}^c r_{\Ccal_i}^*$
\[
    r_\Ccal^* = \#\Scal\left(\frac{r_\Ccal^* \cdot \alpha}{m}: 
   \Scal(r_{\Ccal_1}^*:\Ccal_1)\cup \cdots \cup \Scal(r_{\Ccal_c}^*:\Ccal_c)\right).
\] 
Since the inequality (\ref{ineq:rGcal}) ensures that p-values that were not selected under the condition~(\ref{eq:alg2}) 
 would have not been selected under the condition~(\ref{eq:all}), we conclude that $r_\Ccal^* = r_\Ccal$. 
\end{pf}

\begin{thm}\label{thm:cmp_Om2}
 The FastLSU Algorithm~\ref{alg:2chunks} requires $O(m)$ time and $O(m)$ space where $m$ is the number of p-values to be considered.
The practical memory usage may be restricted to an arbitrary limit for the largest chunk size, $m*$.
\end{thm}
 Proof for the running time and space limitations are shown in the Appendix~\ref{app:proofs}.

\section{Useful tips for finalizing the report of significant p-values}\label{sec:app_tips}
If we assume that the $R$ tests were declared significant by applying either the LSU or FastLSU, we offer tips for finalizing the set of significant p-values. 
 First it is possible to put a bound over the expected value of significant results when applying LSU.
Suppose the set of p-values of interest consists of $m_0$ true null and $m - m_0$, then $E(R) \le \alpha m_0 + m_1$. In the remainder of this section will explain how to protect the $FDR$ control of FastLSU against dependency structures and when such correction is actually needed. We will also explain how to compute q-values (adjusted p-values) for the final results without keeping the entire set of p-values and show how to add a set of simultaneous confidence intervals for the significant test statistics while accounting for the selection effect. 
 
\subsection{Correcting against general case of dependence}\label{sec:dependence}
The LSU procedure is conservative under the general type of positive regression on subsets(PRDS)~\cite{BenjaminiYekutieli2001}, 
so applying the LSU at significance level of $\alpha$ always ensures $FDR\le \alpha$ for PRDS p-values. 
The PRDS class contains a larger set of structures, among them, the independent case and any positively associated p-values such as p-values obtained from a two-sided $t$ tests. 
Since the LSU and FastLSU are equivalent, applying the FastLSU at level $\alpha$ on PRDS p-values will control the FDR at level-$\alpha$, as well. 

For other types of non PRDS dependency, such as in the case of pairwise comparisons, 
it is recommended to use the Benjamini-Yekutieli procedure~\cite{BenjaminiYekutieli2001} that applies LSU using $\alpha^* = \frac{\alpha}{c}$ instead of $\alpha$ where $c=\displaystyle \sum_{k=1}^m\frac{1}{k} \approx \ln m + 0.5772$, where $0.5772$ is the \textit{Euler-Mascheroni constant}. 
By the same argument the FastLSU under non PRDS will have 
$FDR \le \alpha$ when applied under $\alpha^* = \frac{\alpha}{\ln m + 0.5772}$. 
As a matter of practice, we suggest to first perform FastLSU using the significance level $\alpha$. 
Then, if the p-values are non PRDS, correct the $R$ selected p-values by applying FastLSU again over 
the single batch consisting of $m^*=R$ p-values using $\alpha^{**} = \frac{R\alpha^*}{m}$. 
This approach provides $FDR \le \alpha$ since $\alpha^{*}<\alpha$ for $m \ge 2$.

 \subsection{How to compute q-values to a selected subset of significant tests}\label{sec:qvalues}
When it is preferable to report q-values or adjusted p-values, we suggest how this may be done more efficiently. 
If we assume that $R$ tests were selected as significant, let the p-value $p_{(R)}$  be the largest p-value satisfying $p_{(R)} m/R <\alpha$. All p-values larger than $p_{(R)}$ were not selected as significant since $p_{(i)} m/i > \alpha$ for $i>R$ and have q-values $>\alpha$. Therefore it is sufficient to consider only the set of selected $R$ p-values.
 The q-value for the p-value, $p_{(i)}$, under the LSU has the form~\cite{YekutieliBenjamini1999} 
\begin{equation}\label{eq:q-value}
  q_{(R)} = p_{(R)} m/R\quad \mbox{ and }
    q_{(i)} = \min_{j = R,R-1 \ldots, i+1}\{p_{(i)}m/i,\, q_{(j)}\}, \quad \mbox{ for }i<R.
\end{equation}
From this we can see that the algorithms presented by Yekutieli and Benjamini~\cite{YekutieliBenjamini1999} and Storey~\cite{Storey2002} are $O(R \log R)$. 
One needs only sort the $R$ selected p-values in descending order and then
beginning from largest p-value assign the corresponding q-value in a final linear scan recording the minimum q-value assigned thus far. 
The q-values will also be descending assigned in this way and there is no need to compare previous values except
the minimum q-value thus far. The value for the q-value for $q_{(i)}$ will only change when it is less than the minimum seen thus far and the minimum q-values will span a range of p-values until either a sufficient number of p-values have been covered or a sufficient range
in p-values have been covered. Alternatively, one can use existing procedures(such as p.adjust in R software or proc multitest in SAS) to compute adjusted p-values for the set of $R$ selected p-values and multiply the results by $R/m$ to correct them.
  
 \subsection{Confidence intervals for selected subset of significant results.}\label{sec:FCRCIs}

A less common approach in genomic studies is to report the significant test results by constructing a set of confidence intervals for the tests statistics. 
While p-value is a merely measure of the magnitude of the test statistic, a confidence interval may offer the additional information about the dispersion of that magnitude.
The selection adjusted confidence intervals of Benjamini and Yekutieli~\cite{BenjaminiYekutieli2005} offer an appropriate construction that corrects against the false coverage effect of selection.
For a useful example see Jung~\textit{et al}~\cite{JungEtAl2011} usage for the significant log-fold changes of RNA Microarrays.

\section{Discussion}\label{sec:discussion}
We presented an efficient algorithm to apply correctly the Benjamini-Hochberg Linear Step Up $FDR$ controlling procedure in a huge-scale testing problem. 
Since we have shown that our algorithm requires only linear time, 
we can claim that is provably not any more computationally burdensome than even using a rigid Bonferroni cut-off for control.
 However, unlike the rigid Bonferroni cut-off, our approach ensures the $FDR$ control at
 level $\alpha$ and this is a more powerful alternative to controlling the $FWER$, especially when the multiple testing problem is of huge-scale. 
Our approach is also scalable to any subsetting or chunking of the overall subset of p-values. 
 
In addition we offered tips for computational efficiency while performing the LSU over a huge-scale multiple hypotheses testing problem, 
such as showing how to correct for dependency or how to compute q-values directly from the subset of LSU significant results.
We hope that these algorithms and tips offer a better insight into the huge-scale testing problem rather just a black-box of solutions. 
We encourage altering the steps in performing Algorithm~\ref{alg:2chunks}, for instance,  either by applying it sequentially or in
parallel or a mixture of both.

We were, indeed, surprised by the amount of inflated type I error we observed during the exercise on different chunk sizes of the HapMap populations. 
This strongly suggests the need for greater diligence in correctly separating hypotheses tests, whether the objective is to control the $FWER$, the $FDR$, or even Efron-Storey's $pFDR$. 
This observations was also reported by Efron~\cite{Efron2008}, but a full investigation on real data with decreasing chunk sizes was not performed. 
Our suggested approach solves this for the case of the Benjamini-Hochberg $FDR$. 
As we have alluded a similar approach can be adopted to control correctly the Efron-Storey's $pFDR$ and this remains open for further research.
In addition the severe phenomena of inflated type I error we observed suggests more care may be required in reporting and interpreting results
 in genetics literature especially in the case of GWAS and eQTL studies.
Namely to properly understand the significance results there is a need for consistent consideration 
of the algorithms or software used for controlling and separating the hypotheses tests and for record keeping of the chunk sizes
used for a study.

\appendix{}
\section{Proofs and codes}\label{app:proofs}
 \subsection{Proof of Theorem~\ref{thm:cmp_Om}}\label{sec:proof2}
 \begin{pf}[Proof of Theorem~\ref{thm:cmp_Om}]
We present the algorithm here in three steps with running time for each step and accompanying pseudocode. 
\begin{enumerate}
\item[\textbf{Bin.}] Classify all p-values into the bins of the interval, $[0,\alpha]$ 
each of size $\frac{1}{m}$ and keep a total of all p-values with value less than $\alpha$, $m^*$. Label each p-value with a
value $k$ such that $k = \left \lceil{\frac{mp^*}{\alpha}} \right \rceil$ or $k=k+1$ if $ \frac{mp^*}{\alpha} $ equals $ \left \lceil{\frac{mp^*}{\alpha}} \right \rceil$
where $\alpha$ is the significance level and $p^*$ is the given p-value being labeled.  If $p^*$ is labeled with bin $k$ less than or equal to $m$, increment
the count for bin $k$ and increment current p-value count, $m^*$. If a $p^*$ has label $k$ greater than $m$, it may be filtered, so no counts need to be incremented for such p-values (although they can
be labeled with arbitrarily large values for $k$). 
\textit{Running time and memory:} Labeling each p-value and incrementing the labeled bin count requires only
constant time and a single pass through the p-values. In addition to storing the p-values, the p-value labels and bin counts also need to be stored, also
requiring $O(m)$ space each, and a variable for the current p-value count, $m^*$.
\item[\textbf{Accumulate.}] In this step we find the $r^*$ the significant bin to return all p-values in bins  less than or equal to this bin as significant.
To do so, starting from the highest labeled bin's count, i.e,. for $m$, keep a partial sum of the total number of p-values in the bins thus far. If the
current bin's has a non-zero bin count and its value is equal to $m^*$ minus then the current partial sum, then return the current bin as $r^*$ the significant bin.
\textit{Running time and memory}: This step can be done in a single scan of the bin counts and only requires additional variables for the
significant bin, $r^*$, and the partial sum.
\item[\textbf{Return.}] Return as significant all the p-values that were labeled with a $k$ less than or equal to $r^*$.
\textit{Running time and memory} This requires only a single scan of the p-value labels and no additional space.
\end{enumerate}
\end{pf}

\subsection{Pseudocode for Proof of Theorem~\ref{thm:cmp_Om}}\label{sec:Psdcode}
\scriptsize
\begin{verbatim}
Let p_vals[m] be an array of p-values,
p_val_labels[m] be array of bin labels for p-values,
count_bins[min] be an array of bin counts,
m the total number of p-values, and alpha the significance level.
Initialize p_val_labels and count_bins to 0.
sig_pvals = 0; #number of p-vals with small enough significance
for i = 1 to m {
   p* = p_vals[i];
   k = ceil(mp*/alpha); #Bin
   if (  ceil(mp*/alpha) equals mp*/alpha) increment k;
   p_val_labels[i] = k;
   if (k less than or equal to m) {
     increment count_bins[k];
     increment sig_pvals;
   }
}
r* = 0; #significant bin
total_p_vals = 0; # partial sum of p_values
for i = m to 1 {
  add count_bins[i] to total_p_vals]; #Accumulate
  if (sig_pvals - total_p_vals equals i) {
       r* = i and break;
  }
}
for i = 1 to m {
     if (p_val_labels[i] less than or equal to r*) #Return
        return p_vals[i] as signficant;
}
\end{verbatim}
\normalsize

\subsection{Example for Algorithm~\ref{alg:2chunks}}\label{exe:3}
\begin{exmp}\label{ex:exmp3} To demonstrate the Algorithm~\ref{alg:2chunks} consider the $15$ p-values appearing in Example~\ref{ex:exmp1} and also in the original LSU paper~\cite{BenjaminiHochberg1995}. 
Further assume that for some reason the $15$ p-values are divided into two chunks. The first chunk, say $\Ccal_1$, consists of the first $8$ p-values:
\[
\Ccal_1 = \{0.6528,0.7590,0.0298,0.4262,0.0459,0.0278,0.0001,0.0019\},
\]
and the second chunk, $\Ccal_2$
\[
\Ccal_2 = \{0.0004,0.0201,1.0000,0.5719,0.3240,0.0095,0.0344\}.
\]
 Algorithm~\ref{alg:2chunks} applied for $\alpha = 0.05$ has three steps:
 In Step 1, we scan the $8$ p-values in $\Ccal_1$ and seek for the largest p-value, $p_{(i:\Ccal_1)}$ satisfying $<(i+15-8)\alpha/15$. This can be done in a similar manner to applying Algorithm~\ref{alg:1chunk}. First scan gives $5$ p-values $<(8+15-8)0.05/15=0.05$. Second scan gives $4$ p-values $<(5+15-8)0.05/15=0.04$ and stops with the $4$ p-values $<(4+15-8)0.05/15=0.0367$. The selected p-values for the first step are 
 \[
\Scal(0.0367:\Ccal_1) = \{0.0298,0.0278,0.0001,0.0019\},
\]
In Step 2, we scan the $7$ p-values  in $\Ccal_2$ and seek for the largest $p_{(i:\Ccal_2)}<(i+15-7)0.05/15$. First scan gives $4$ p-values $<(7+15-7)0,05/15=0.05$ and immediately stops with these $4$ p-values $<(4+15-7)0.05/15=0.04$. The selected p-values are
 \[
\Scal(0.04:\Ccal_2) =\{0.0004,0.0201,0.0095,0.0344\}.
\]
Next, we follow the last step of Algorithm~\ref{alg:2chunks} which is a combination step and applies on the collection of $8$ p-values selected at the former steps: 
 \[
\Scal(0.0367:\Ccal_1) \cup \Scal(0.04:\Ccal_2) = \{0.0298,0.0278,0.0001,0.0019,0.0004,0.0201,0.0095,0.0344\}.
\]
All $8$ selected p-values are clearly $<0.05$. Second, $5$ p-values$<8\cdot 0.05/15=0.0267$,and third scan finds out $4$ p-values $<5\cdot 0.05/15=0.0167$ that also satisfy $<4\cdot 0.05/15$.
The selected p-values are, again, $0.0001,0.0019,0.0004,0.0095$.
\end{exmp}
\subsection{Proof of Theorem~\ref{thm:cmp_Om2}}\label{sec:proof}
 \begin{pf}[Proof of Theorem~\ref{thm:cmp_Om2}.]
For $m$ p-values arbitrarily divided into $n$ chunks of size $m_c$ for $c=1,...,n$ such that the maximum chunk size is $m^{*}$,
we show that the algorithm is still linear in $m$ and never uses more than $m^{*}$ space.
\begin{enumerate}
\item[\textbf{Bin.}] This step is as for Algorithm~\ref{alg:1chunk}. It is important to note that the binning is done relative to $m$ and not the size of the chunk.
The only important difference is that bin sums are not maintained because of the $m^{*}$ space limitation. Labels need not be stored either.
This step also counts the sum, $m'$, of all p-values across all groups that are less than $\alpha$.
\textit{Running time and memory:} This requires a linear scan. A count variable can be kept for each group in order to get $m'$.
\item[\textbf{Accumulate.}] For each group, find the bin sums for the largest $m^{*}$ partition of bins not covered yet, i.e., find bin
sums for bins $m'-jm^{*}+1$ to $m-(j-1)m^{*}$ for $j=1,...,n$ and we do both of the following before incrementing $j$. Accumulate
bin sums across the chunks. (This can be done in a linear scan of the chunks and a single array accumulating bin sums across chunks.)
Finding $r*$ is as in Algorithm~\ref{alg:1chunk} Step 2. However, now only $m^{*}$ bins may be checked at a pass before needed to increment $j$. The
subtotal of p-values counted thus far is maintained after $j$ is incremented.
\textit{Running time and memory}: This step must be repeated at most $n$ times and requires a linear scan of the data.
At any point at most $m^{*}$ space plus several count variables are used. 
\item[\textbf{Return.}] This step is as in Algorithm~\ref{alg:1chunk}.
\end{enumerate}
\end{pf}

\subsection{R script for the FastLSU with examples}\label{sec:RScript}
\scriptsize
\begin{verbatim}
fast.LSU = function(alpha=0.1,pvls,mtc,mtg){
# alpha: the desired significance level;
# mtg: the global number of p-values in the problem (the sum of all p-values in all chunks); 
# mcg: total number of p-values in the chunk (<=mtg);
# pvls: pvalues;
# -------------------------------------------------------------------------------------; 
   m = length(pvls); #no of actual p-values at the input chunk (might be < mtc, e.g., step c+1 in algorithm 2);
   print(paste('FastLSU for group of',mtg,'p-values.'))
   if(m<mtc){
     print(paste(m,'p-values were detected at the input chunk. Original chunk size is assumed to be',mtc));
     print('Make sure your chunk size is correct.');
     print('Input chunk size can smaller than the assumed chunk size at the final step c+1 of Algorithm 2.');
   } 
   if(mtc<mtg){print(paste('Input chunk size is',mtc,'. Group size is', mtg,'p-values.'))} 
   sig    = rep(1,m); #flag, for significance, 1 is default;
   r.new  = m; 
   r.old  = m-1; 
   diff.r = 1; 
   stp = 0; #follow steps;
   while(diff.r>0){
        stp = stp+1
        sig[pvls>=(r.new+mtg-mtc)*alpha/mtg] = 0; #set the sig flag of large enough p-values to zero;
        r.old = r.new; 
        r.new = sum(sig);
       print(paste('On step',stp,':',sum(sig),'p-values were selected for candidance of significant.'))
      diff.r = r.old - r.new
  }
  return(list(pvalues.c=pvls[sig==1],r=r.new,sig=sig))
} 

#simulate p-values for example;
pvals.sim = runif(30000); # random U[0,1] (null) p-values;
Bsig = 1-rbinom(30000,1, .02); Bsig[Bsig==0] = .0001;
pvals.sim = pvals.sim*Bsig; #add 2% of small p-values to be significant;

# Example of a single chunk;
results.allchunks = fast.LSU(alpha=0.1,pvls=pvals.sim,mtc=length(pvals.sim),mtg=length(pvals.sim));

# Example of 2 chunks of p-values (1st of 10,000, 2nd of 20,000);
# chunking;
pv.chunk1 = pvals.sim[1:10000];
pv.chunk2 = pvals.sim[10001:30000];

# Step 1:
results.chunk1 = fast.LSU(alpha=0.1,pvls=pv.chunk1,mtc=length(pv.chunk1),mtg=length(pvals.sim));

# Step 2:
results.chunk2 = fast.LSU(alpha=0.1,pvls=pv.chunk2,mtc=length(pv.chunk2),mtg=length(pvals.sim));

# Step 3: # get final candidancy; 
cand.pvals = t(cbind(t(results.chunk1$pvalues.c),t(results.chunk2$pvalues.c)));
# keep mtc equal to mtg, the fast.LSU function already considers length(cand.pvals).
results.final = fast.LSU(alpha=0.1,pvls=cand.pvals,mtc=length(pvals.sim),mtg=length(pvals.sim));
\end{verbatim}
\normalsize

\subsection{SAS Macro}\label{sec:SASFastLSU}
\scriptsize
\begin{verbatim}
%macro calc_T_BH(alpha=0.1,pvaldata,pvals,mtg,mtc);
* pvaldata: the name of SAS data set that contains the p-values;
* pvalue: the SAS variable name for the p-values at the SAS data set;
* alpha: the desired significance level;
* mtg: the global number of p-values in the problem; 
* mcg: the total number of p-values in the chunk;

data ctemp; set &pvaldata.;  
* m is the actual number of p-values at the input chunk (m <= mtc);
* For example: see Step c+1 at Algorithm 2 (m is the number of p-values canditates);
call symput("m",_N_); run;

*set starting values;
%let rtcest = &m.;
%let rold   = %eval(&rtcest.+1);
%let k      = 0;
data ctemp; set ctemp; if &pvalue. < &alpha.; run;

%do %while(&rtcest.< &rold.);
    %let k    = %eval(&k.+1);
    %let rold = &rtcest.; 
    *apply the steps of fastLSU;
   data ctemp; set ctemp; 
      if &pvalue. < (&rtcest.+ &mtg.-&mtc.) / &mtg. * &alpha.; 
   run;

* recompute mtc estimate for fast BH;
data _NULL_; set ctemp; call symput("rtcest",_N_); run;
%let rtcest=&rtcest.;
%end; *end while loop;
data sig; set ctemp; run;
%mend;

%calc_T_BH(alpha=0.1,pvaldata=StrangerCEU,pvalue=pvalue,mtg=13755172,mcg=13755172);
\end{verbatim}
\normalsize


\end{document}